\documentclass[aps,twocolumn,nofootinbib]{revtex4}

\usepackage{amsmath,amssymb}
\usepackage{mathrsfs}
\usepackage{bm}
\usepackage{graphicx}
\usepackage[hidelinks]{hyperref}
\usepackage{comment}
\usepackage{subcaption} 
\usepackage[dvipsnames]{xcolor}
\usepackage{enumitem} 
\usepackage{physics}
\usepackage[normalem]{ulem}
\usepackage{bibunits}
\usepackage{mhchem}
\defaultbibliography{bibliography}      
\defaultbibliographystyle{apsrev4-2}  

\DeclareMathOperator{\St}{\mathbb{S}}

\newcommand{\n}{\mathbf{n}}
\newcommand{\y}{\mathbf{y}}
\newcommand{\Y}{\mathbf{Y}}

\newcommand{\N}{\mathbf{N}}
\newcommand{\V}{\mathrm{V}}
\newcommand{\Om}{\boldsymbol{\Omega}}

\newcommand{\g}{\mathbf{g}}
\newcommand{\p}{\mathbf{p}}
\newcommand{\x}{\mathbf{x}}

\newcommand{\X}{\mathbf{X}}

\newcommand{\w}{\mathbf{w}}
\newcommand{\jj}{\mathbf{j}}

\newcommand{\pp}{\mathfrak{p}}
\newcommand{\jjj}{\mathbf{j}}
\newcommand{\sj}{j}
\newcommand{\J}{\boldsymbol{J}}

\newcommand{\z}{\zeta}

\newcommand{\0}{\mathbf{0}}

\newcommand\wh[1]{\widehat{#1}}

\newcommand\mcl[1]{\mathcal #1} 
\newcommand\mbb[1]{\mathbb #1}

\newcommand{\pars}[1]{\left( #1 \right)}

\usepackage{hyperref}
\hypersetup{%
 setpagesize=false,%
 bookmarks=true,%
 bookmarksdepth=tocdepth,%
 bookmarksnumbered=true,%
 colorlinks=false,%
 pdftitle={},%
 pdfsubject={},%
 pdfauthor={},%
 pdfkeywords={}}

\newcommand{\Real}{\mathbb{R}}
\newcommand{\Integer}{\mathbb{Z}}
\usepackage{ifthen}
\newboolean{showcomments}
\setboolean{showcomments}{true}
\newcommand{\COMM}[1]{%
  \ifthenelse{\boolean{showcomments}}{%
    {\textcolor{red}{\small [TJK: #1]}}%
  }{}%
}
\newcommand{\COMMIM}[1]{%
  \ifthenelse{\boolean{showcomments}}{%
    {\textcolor{teal}{\small [IM: #1]}}%
  }{}%
}

\begin{document}


\title{Large-deviations theory for growing chemical reaction networks}

\author{Praful Gagrani}
\affiliation{Institute of Industrial Science, The University of Tokyo, Tokyo 153-8505, Japan}

\author{Ignacio Madrid}
\affiliation{Institute of Industrial Science, The University of Tokyo, Tokyo 153-8505, Japan}

\author{Tetsuya J Kobayashi}
\affiliation{Institute of Industrial Science, The University of Tokyo, Tokyo 153-8505, Japan}

\date{\today}

\begin{abstract}
Growing biochemical systems are intrinsically noisy, with one important source
of stochasticity arising from the discrete reaction events of the underlying
biochemical network.  Because growing systems do not generally admit
stationary abundance distributions, it is unclear how to separate this
intrinsic chemical noise from other sources of variability in population and
single-cell data.  Here we develop a large-deviation theory for exponentially
growing stochastic chemical reaction networks by decomposing abundance into
volume and composition.  In these coordinates, balanced growth corresponds to
a stable composition together with exponentially increasing volume.  We show
that composition fluctuations satisfy a large-deviation principle with speed
equal to the volume of the growing system, and the corresponding quasipotential is selected by a solution of the contact Hamilton--Jacobi equation.  We also derive a fluctuation
theory for accumulated observables and show that their covariances decay with
accumulated-volume, rather than with physical time.  Applications to a minimal
autocatalytic network and a coarse-grained cellular growth model demonstrate
how the theory predicts composition quasipotentials, growth-rate fluctuations,
and correlations between observables.  The framework therefore provides a
stochastic null model for single-cell heterogeneity generated by the intrinsic
stochasticity of the underlying chemical reaction network.
\end{abstract}


\maketitle

\section{Introduction}

Metabolism in growing organisms is often modeled using chemical reaction
networks (CRNs).  At the cellular scale, the underlying reaction events are
stochastic, and this intrinsic stochasticity propagates to experimentally
accessible quantities such as molecular composition, growth rate, and
time-averaged proteome fractions.  A defining feature of balanced growth,
however, is the coexistence of extensive growth and intensive stability:
molecular abundances increase in time while quantities such as composition
remain approximately stationary.  The aim of this work is to develop a
large-deviation and fluctuation theory for stochastic CRNs in this growing
regime.

Large-deviation theory characterizes the probabilities of rare fluctuations
and provides a systematic description of fluctuations around typical
behavior \cite{FengKurtz2006,Touchette2009}.  Stochastic CRNs form an
important class of continuous-time Markov jump processes.  For closed or
confined CRNs admitting a stationary state, the large-system-size theory is
naturally organized around the stationary distribution.  In the limit of a
large system-size parameter \(K\), this distribution often takes the
large-deviation form
\[
    \pi^K(\y)
    \asymp
    \exp[-K\mcl V(\y)],
\]
where \(\mcl V\) is the stationary rate function, commonly referred to as the
quasipotential, and \(K\) sets the large-deviation speed
\cite{anderson2015lyapunov,snarski2021hamilton,AgazziDemboEckmann2018}.
The quasipotential plays the role of a nonequilibrium free energy: its minima
identify stable deterministic attractors, its differences determine
exponential scales for rare transitions, and it can be characterized through
a stationary Hamilton--Jacobi equation.

A complementary large-deviation theory describes fluctuations of
time-integrated observables on trajectory space.  For an ergodic Markov
process, the probability that a time-averaged observable takes an atypical
value typically scales as \cite{eyink1996action}
\[
    \mbb P(\widehat O_t\approx o)
    \asymp
    e^{-t I(o)},
\]
so that physical time \(t\) plays the role of the large-deviation speed.
The corresponding scaled cumulant generating function is obtained from a tilted generator, and the Markov process conditioned on an atypical value of
the observable can be represented by a generalized Doob transform
\cite{chetrite2013nonequilibrium,chetrite2015nonequilibrium, chetrite2015variational}.  These ideas have been extended to stochastic CRNs
in the large-volume limit, where the tilted problem acquires a
Hamiltonian--Lagrangian description and Hamilton--Jacobi equations determine
the effective conditioned dynamics
\cite{lazarescu2019large,chabane2021rarity,chabane2022effective}.
They show that the cumulant generating function (CGF) for a time-averaged observable is proportional to $Kt$, which consequently means that the variance of a time-averaged observable scales as \((Kt)^{-1}\).

In this work, we focus on growing CRNs of the type commonly used to model
bacterial growth, for which the deterministic dynamics admit exponentially
growing attractors
\cite{pandey2016analytic,pandey2020exponential,lin20,muller2022elementary}.
Neither of the standard asymptotic pictures for stationary CRNs applies
directly in this setting.  In a supercritical growing CRN, the abundance
process does not remain in a compact region of state space and therefore does
not admit a stationary quasipotential in abundance coordinates.  At the same
time, the reaction activity is itself nonstationary: as the system grows, the
number of reaction events occurring per unit physical time increases with its
volume.  Consequently, neither the fixed system-size scale \(K\) nor physical
time \(t\) alone captures the asymptotic fluctuations.  The central question
of this work is therefore what replaces these two scales in an exponentially
growing reaction network.

Motivated by the standard analysis of deterministic growth dynamics, we
decompose the abundance as \(\y=v\x\), where \(v\) is the volume and
\(\x\) is the composition.  The composition evolves on the simplex
\(\Delta_\w=\{\x\geq0:\w^\top\x=1\}\), while the volume evolves
multiplicatively.  In balanced growth, the volume increases exponentially
while the composition approaches a stable fixed point.  Our first main result
is that, within the growing basin of a balanced-growth state \(\x^\star\),
composition fluctuations obey the large-deviation form
\[
    \mbb P(\X_t^K\approx\x)
    \asymp
    \exp[-K v_t\,\mcl I(\x)],
\]
where \(v_t\) is the volume at which the composition is observed and
\(\mcl I(\x)\) is the composition quasipotential.  Thus the effective
large-deviation speed is the instantaneous unnormalized volume \(K v_t\), rather
than the fixed scale \(K\).  We show that \(\mcl I\) is characterized by a
stationary contact Hamilton--Jacobi equation.  The contact structure arises
because the volume coordinate has been eliminated from the state description
but remains in the action; its effect survives in the composition dynamics
through an additional dissipative term in the characteristic equations.

Our second main result concerns fluctuations of long-time observables.  Since
reaction activity is proportional to the current volume, the natural clock for
such statistics is not physical time but the accumulated-volume,
\[
    A_t^K=\int_0^t V_s^K\,ds.
\]
We therefore introduce accumulated-volume-averaged estimators, which have the
same balanced-growth limits as their conventional time-averaged counterparts
but different fluctuation statistics.  Using a tilted Hamilton--Jacobi
problem, we derive their scaled cumulant generating function and show that
their covariances decay on the accumulated-volume scale,
\[
    \mathrm{Cov}\!\left(\widehat{\mcl O}_t^K\right)
    \sim
    \left[
        K v_0\frac{e^{\Lambda t}-1}{\Lambda}
    \right]^{-1},
\]
up to the covariance matrix determined by the Hessian of the scaled cumulant
generating function.  Here \(\Lambda\) is the deterministic exponential growth
rate.  The conventional constant-volume scaling is recovered continuously in
the limit \(\Lambda\to0\), since
\(
(e^{\Lambda t}-1)/\Lambda\to t
\).

The paper is organized as follows.  In Sec.~\ref{sec:setup}, we introduce
stochastic CRNs, composition--volume coordinates, and the
accumulated-volume observables used throughout the paper.  We state the two
main results in Sec.~\ref{sec:results}, while their derivation is deferred to
Appendix.  In Sec.~\ref{sec:applications}, we apply the
framework to a minimal autocatalytic network and to a coarse-grained model of
cellular growth with transport machinery, precursor, and ribosomes.  We
conclude in Sec.~\ref{sec:discussion} by discussing the implications of the
framework for experiments and outlining directions for future work.

\begin{table*}[t]
\caption{
Summary of quantities used and their symbols in the stochastic and deterministic descriptions.
}
\label{tab:variables_summary}
\begin{ruledtabular}
\begin{tabular}{p{0.18\textwidth}p{0.25\textwidth}p{0.24\textwidth}p{0.24\textwidth}}
Quantity
&
Meaning
&
Stochastic symbol
&
Deterministic symbol
\\
\hline
Copy-number
&
Molecular counts of all species.
&
\(\N_t^K=(N_{1,t}^K,\ldots,N_{D,t}^K)\)
&
\(K\y(t)=K V(t)\x(t)\)
\\[0.6em]

Abundance
&
Counts scaled by the system-scale parameter.
&
\(\Y_t^K=\N_t^K/K\)
&
\(\y(t)\)
\\[0.6em]

Composition
&
Intensive state obtained by quotienting out volume.
&
\(\X_t^K=\Y_t^K/V_t^K\)
&
\(\x(t)=\y(t)/v(t)\)
\\[0.6em]

Normalized volume
&
Extensive coordinate in abundance units.
&
\(V_t^K=\w^\top\Y_t^K\)
&
\(v(t)=\w^\top\y(t)\)
\\[0.6em]

Unnormalized volume
&
Total volume scale in copy-number units.
&
\(K V_t^K=\w^\top\N_t^K\)
&
\(K v(t)\)

\\[0.6em]
Growth rate
&
Asymptotic exponential growth rate of det.\ abundance.
&
\qquad ---
&
\(\Lambda\)

\\[0.6em]
Accumulated-volume
&
Integral of the normalized volume along growth.
& 
\( A_t^K = \int_0^t V^K_s \,ds\)
&
\( a_t^* = v_0 \int_0^t e^{\Lambda s}\,ds = \frac{v_0(e^{\Lambda t}-1)}{\Lambda}\)
\end{tabular}
\end{ruledtabular}
\end{table*}

\section{Setup }
\label{sec:setup}
This section introduces the stochastic variables, deterministic limits, and
observables used throughout the paper. Table~\ref{tab:variables_summary} summarizes
the notation introduced in this section.

\subsection{Stochastic and deterministic chemical reaction networks}

Consider a stochastic chemical reaction network (CRN) with \(D\) species and \(R\) reactions.  Let \(\N_t^K\in \mathbb N^D\) denote the vector of species copy-number at time \(t\), where \(K\in \Real_{>0}\) is a parameter setting the system-scale, such that $\N_0^K = O(K)$. We suppose that $\N_t^K$ evolves according to the stochastic dynamics described by
\begin{equation}
    \N^K_t
    =
    \N_0^K
    +
    \sum_{r=1}^{R} \St^r \Om_t^{K,r},
    \nonumber
\end{equation}
where
\(\St^r\in\mathbb \Integer^S\) is the stoichiometric change vector of reaction \(r\) and $\Om_t^{K,r}\in \mathbb N$ is the cumulative reaction count number of reaction \(r\) until time $t$. We suppose that the counting processes $\Om_t^{K,r}$ are written as
\begin{equation}
    \Om_t^{K,r}
    =
    \mcl{P}^r\pars{\int_0^t \J_r^K(\N_s^K)\,ds}, \nonumber
\end{equation}
where \( (\mcl{P}^r(t))_{t \geq 0}, \, r = 1,...,R,\) are $R$ independent unit-rate Poisson processes and \(\J_r(\n)\in \Real\) is the propensity function of reaction \(r\). See \cite{anderson2015stochastic} for a detailed construction of these counting processes. 

We introduce the abundance coordinate
\begin{equation}
    \Y_t^K
    =
    \frac{\N_t^K}{K}\in \Real_{\ge 0}^{D}.
\end{equation}
For instance, \(K\) may be chosen to be Avogadro's number, in which case \(\Y_t^K\) is measured in moles.  We assume that the propensity functions satisfy the scaling
\begin{equation}
    \lim_{K\to\infty}
    \frac{\J^K_r(K\,\y)}{K}
    =
    j_r(\y),\nonumber
\end{equation}
where the limit function $\jj : \Real_{\ge 0}^{D} \to \Real_{\ge 0}^{R}$ is called the deterministic rate function. Then, in the limit \(K\to\infty\), under standard regularity assumptions, the stochastic process $\Y_t^K$ initialized at $\Y_0^K =  \y_0$  converges to the deterministic CRN dynamics (see \cite{ethierkurtz}):
\begin{equation}
    \dot \y
    =
    \sum_{r=1}^R \St^r j_r(\y), \qquad \y(0) = \y_0.
    \label{eq:concentration_dynamics}
\end{equation}

\subsection{Composition--volume (CV) coordinates}
We introduce a positive vector
\(\w\in\mathbb R^D_{\geq 0}\) that quantifies the physical volume of each species. We assume that the total unnormalized physical volume of the system is linear in the copy-number and equals $\w^\top \N_t^K$.
We define the normalized volume (and simply ``volume" from now one) as the rescaled quantity 
\begin{equation}
    V_t^K := \frac{1}{K} \w^\top \N_t^K =  \w^\top \Y_t^K.
\end{equation}
Indeed, if the vector \(\w\) carries physical units of volume per unit molecule, and $K$ is chosen to be the Avogadro's number, \(V\) represents the molar volume of the system.

We define the composition variable as
\begin{equation}
    \X_t^K
    :=
    \frac{\N_t^K}{\w^\top \N_t^K} = \frac{\Y_t^K}{\V_t^K},
    \label{eq:stochastic_composition_process}
\end{equation}
and thus $\N_t^K = K \Y_t^K = K V_t^K \X_t^K$.  Since \( \w^\top \X_t^K=1\) holds,  \(\X_t^K\) lies on the \((D-1)\)-dimensional simplex~
\[
    \Delta_\w
    :=
    \left\{
    \x\in\mathbb R_{\ge 0}^S:
    \w^\top \x=1
    \right\}.
\] 
The coupled processes $(\V_t^K,\X_t^K)$ and $K$ completely determine $\N_t^K$. We work in this coordinate system throughout the paper.

\subsection{Deterministic composition dynamics}
We assume that the deterministic reaction rates \(j_r\) are homogenous functions of degree one \cite{pandey2020exponential} (or scalable, \cite{lin20}), that is, 
for all \(\x\in\Delta_{\w}\) and \(v>0\), we can write
\begin{equation}
    \frac{j_r(v\x)}{v}
    =
    \sj_r(\x),
    \qquad r=1,\ldots,R.
    \label{eq:extensive_kinetics}
\end{equation}

Let
\begin{equation}
    \nu_r
    :=
    \w^\top \St^r 
    \label{eq:volume_stoichiometric_change}
\end{equation}
denote the volume change associated with reaction \(r\).  

We call $(v, \x)$ the deterministic limits as $K \to \infty$ of the couple $(\V^K,\X^K)$. Starting from the limit ODE in Eq.~\eqref{eq:concentration_dynamics} and substituting \(\y=v \x\), we see that $(v, \x)$ is solution of the system
\begin{align}
    \dot v
    &=
    \w^\top \dot \y
    =
    \sum_{r=1}^R \nu_r j_r(v\x)
    =
    v\sum_{r=1}^R \nu_r \sj_r(\x),
    \label{eq:volume_dynamics}
    \\
    \dot \x
    &=
    \frac{\dot \y}{v}
    -
    \x\frac{\dot v}{v}
    =
    \sum_{r=1}^R \St^r \sj_r(\x)
    -
    \x\sum_{r=1}^R \nu_r \sj_r(\x).
    \label{eq:composition_dynamics}
\end{align}
Thus the composition dynamics is closed on the simplex \(\Delta_\w\), while the volume evolves multiplicatively.

We call a balanced-growth state, a stable fixed point $\x^\star$ of the composition dynamics \eqref{eq:composition_dynamics}.  It satisfies
\begin{equation}
    \sum_{r=1}^R \St^r \sj_r(\x^\star)
    =
    \Lambda \x^\star,
\end{equation}
where
\begin{equation}
    \Lambda
    =
    \sum_{r=1}^R \nu_r \sj_r(\x^\star)
    \label{eq:growth_rate}
\end{equation}
is the exponential growth rate. When \(\Lambda>0\), the system grows
exponentially in volume, and if $\x^\star$ is a unique globally stable fixed point, then the composition converges to $\x^\star$ as the system grows.

The volume-dependent kinetics above model non-dilute mixtures.  In Appendix, we show that the usual
dilute, constant-volume approximation is recovered as a limiting case of the above dynamics. Specifically, to obtain the results for conventional dilute systems using the results presented here, one may set \(\nu_r=0\), remove the
solvent from the dynamical variables, and treat the volume as constant.

\subsection{Observables}
\label{sec:observables}

We consider observables measured along individual stochastic trajectories.
There are two natural classes \cite{jack2015effective}.  \emph{State-dependent observables} depend on
the state of the system at a given time, for example a molecular composition.
\emph{Jump-dependent observables} depend on transitions between states, for
example reaction fluxes or changes in volume.  For either class, each
trajectory produces a random estimator; ensemble averages, variances, and
covariances of these estimators then characterize the stochasticity across
realizations.

For growing systems there are also two natural ways of averaging along a
trajectory.  The conventional choice is to average uniformly in physical
time.  We refer to these as \emph{time-averaged} estimators.  Alternatively,
because the number of reaction events per unit time grows proportionally to
the system volume, we can weight observations by the instantaneous volume.
We refer to these as \emph{accumulated-volume-averaged} estimators.  Introducing
the stochastic accumulated-volume
\begin{equation}
    A_t^K
    :=
    \int_0^t V_s^K\,ds,
    \label{eq:setup_integrated_volume}
\end{equation}
we define the two classes of estimators below.  The \(F\)-type and \(G\)-type
observables used throughout the paper will refer specifically to the
accumulated-volume-averaged estimators.

\subsubsection{State-dependent observables}
Let \(f_i:\Delta_\w\rightarrow\Real\) a function of the composition.  Its conventional time-averaged estimator along a trajectory is
\begin{equation}
    \widehat f_{i,t}
    :=
    \frac{1}{t}
    \int_0^t
    f_i(\X_s^K)\,ds .
    \label{eq:state_observable_time_average}
\end{equation}
We instead define the corresponding accumulated-volume-averaged, or
\(F\)-type, estimator by
\begin{equation}
    \widehat F_{i,t}^{K}
    :=
    \frac{
        \displaystyle\int_0^t
        V_s^K f_i(\X_s^K)\,ds
    }{
        \displaystyle\int_0^t V_s^K\,ds
    }
    =
    \frac{1}{A_t^K}
    \int_0^t
    V_s^K f_i(\X_s^K)\,ds ,
    \label{eq:state_observable_integrated_volume}
\end{equation}
where \(i=1,\ldots,m .\)
Thus the time average assigns equal weight to equal intervals of physical
time, whereas the \(F\)-type estimator assigns weight in proportion to the
volume, and hence approximately to the amount of reaction activity occurring
during that interval.

As a simple example, consider the composition fraction
\begin{equation}
    f(\x)
    =
    \frac{x_1}{x_1+x_2}. \nonumber
\end{equation}
The two estimators are
\begin{align*}
    \widehat f_t
    &=
    \frac{1}{t}
    \int_0^t
    \frac{X_{1,s}^K}{X_{1,s}^K+X_{2,s}^K}\,ds ,
    \\
    \widehat F_t^K
    &=
    \frac{
        \displaystyle\int_0^t
        V_s^K
        \frac{X_{1,s}^K}{X_{1,s}^K+X_{2,s}^K}\,ds
    }{
        A_t^K
    }.
    \label{eq:composition_estimators_example}
\end{align*}

\subsubsection{Jump-dependent observables}
The second class assigns a value to changes of state.  Let
\(\Om_t^K=(\Om_t^{K,1},\ldots,\Om_t^{K,R})^\top\) denote the vector of
cumulative reaction counts, and let
\[
    \g_j=(g_{j,1},\ldots,g_{j,R})^\top
\]
assign a contribution \(g_{j,r}\) whenever reaction \(r\) occurs.  More
generally these weights could depend on the state before or after each jump;
here we restrict to constant reaction weights for simplicity.

A conventional time-averaged estimator of the corresponding intensive flux
is
\begin{equation}
    \widehat g_{j,t}
    :=
    \frac{1}{t}
    \int_0^t
    \frac{
        \g_j^\top d\Om_s^K
    }{
        K V_{s^-}^K
    },
    \qquad j=1,\ldots,\ell ,
    \nonumber
\end{equation}
where \(V_{s^-}^K\) is the volume immediately before the jump.  This
normalizes each reaction event by the instantaneous system size and then
averages the resulting flux uniformly in physical time.

The corresponding accumulated-volume-averaged, or \(G\)-type, estimator is
\begin{equation}
    \widehat G_{j,t}^K
    :=
    \frac{
        \g_j^\top\Om_t^K
    }{
        K A_t^K
    }
    =
    \frac{
        \g_j^\top\Om_t^K
    }{
        K\int_0^t V_s^K\,ds
    } .
    \label{eq:change_observable_integrated_volume}
\end{equation}
Rather than normalizing each event separately and subsequently averaging in
time, this estimator divides the total accumulated reaction count by the
total integrated system size over which those reactions occurred.

The growth rate provides an important example.  Taking
\begin{equation}
    g_r=\nu_r=\w^\top\St^r ,\nonumber
\end{equation}
and using the fact that reaction \(r\) changes the volume by
\(\nu_r/K\), we have the exact identity
\begin{equation}
    \frac{1}{K}
    \sum_{r=1}^R
    \nu_r\Om_t^{K,r}
    =
    V_t^K-V_0^K .
    \nonumber
\end{equation}
The time-averaged growth-rate estimator is therefore
\begin{equation}
    \widehat\mu_t
    :=
    \frac{1}{t}
    \sum_{r=1}^R
    \int_0^t
    \frac{\nu_r}{K V_{s^-}^K}
    \,d\Om_s^{K,r}.
    \nonumber
\end{equation}
In the large-system limit this is the stochastic analogue of the time average
of the instantaneous specific growth rate,
\begin{equation}
    \widehat\mu_t
    \simeq
    \frac{1}{t}
    \log\frac{V_t^K}{V_0^K}.
\end{equation}
The \(G\)-type growth-rate estimator instead takes the particularly simple
form
\begin{equation}
    \widehat G_t^{K}
    =
    \frac{
        V_t^K-V_0^K
    }{
        A_t^K
    }
    =
    \frac{
        V_t^K-V_0^K
    }{
        \displaystyle\int_0^t V_s^K\,ds
    }.
    \label{eq:growth_rate_integrated_volume}
\end{equation}

In balanced growth, both averaging schemes estimate the same asymptotic
quantities.  In particular,
\begin{align}
    \mbb{E}[\widehat f_{i,t}],
    \mbb{E}[\widehat F_{i,t}^{K}]
    &\xrightarrow{t,K \to \infty}
    f_i(\x^\star),
    \nonumber\\
    \mbb{E}[\widehat g_{j,t}],
    \mbb{E}[\widehat G_{j,t}^{K}]
    &\xrightarrow{t,K \to \infty}
    \sum_{r=1}^R
    g_{j,r}\sj_r(\x^\star),
    \nonumber
\end{align}
with limits taken first in $K \to \infty$, then $t \to \infty$.  A short argument is given in
Appendix.
Although the two estimators have the same asymptotic means,
their fluctuation statistics are different. Our results concern the accumulated-volume-averaged observables.  We return to
the conventional time-averaged estimators only in the applications, where we
use them as a comparison to highlight the different fluctuation behavior of
the two averaging schemes.

\section{Results}
\label{sec:results}
The central difficulty in growing CRNs is that abundance space is
nonstationary: in the supercritical regime, the total volume grows
exponentially.  The key simplification is to separate abundance into volume
and composition, \(\y=v\x\).  Although the volume grows, the composition may
converge to a balanced-growth fixed point and fluctuate around it.  We show
that these fluctuations satisfy a large-deviation principle with speed
\(K v\), where \(K\) is the system-scale parameter and \(v\) is the observed
volume.  The resulting composition quasipotential is characterized by a
stationary contact Hamilton--Jacobi equation. We then study the fluctuations of accumulated-volume-averaged observables when the system is in the balanced-growth regime. We give
a characterization and an explicit approximation formula for calculating their long-time means and covariances.  

For clarity of exposition, throughout this section and in the examples below
we restrict attention to systems for which
Eq.~\eqref{eq:composition_dynamics} admits a unique globally asymptotically
stable fixed point \(\x^\star\) with balanced-growth rate \(\Lambda>0\).
We further assume that the stochastic process does not become extinct and
reaches every finite target volume and target accumulated volume almost surely.
Under these assumptions, no survival conditioning is required. The more
general theory, which allows for reverse reactions, degradation, death, and
other mechanisms for which the target attainment may fail, is developed in
Appendix, and the results presented here follow as a special case.

\subsection{Composition quasipotential}

Our first main result is that the considered growing CRNs admit a generalized quasipotential on composition space characterized by a contact Hamiltonian (see for example \cite{BravettiCruzTapias2017} for a detailed review on contact Hamiltonian mechanics). 

For \(v>v_0\), define the first time at which the volume attains the value of $v$,
\begin{equation}
    \tau_v^K:=\inf\{t\geq0:V_t^K\geq v\}.    
    \nonumber
\end{equation}
The theory developed in Appendix establishes the large-deviation asymptotic\footnote{
We say $a_n(x) \asymp b_n(x)$ if $\lim_{n \to \infty} \frac{1}{n}\log \frac{a_n(x)}{b_n(x)}=0$ or, equivalently, 
$a_n(x) = b_n(x)\cdot e^{o(n)}$.
}
\begin{equation}
\mbb{P}_{\x_0,v_0}\!\left(\X_{\tau_v^K}^K\approx \x\right)
    \asymp \exp\{-K v\mcl{I}(\x) \},
    \label{eq:composition_ldp}
\end{equation}
with limits taken first in \(K\to\infty\) and second \(v/v_0\to\infty\). Starting from some volume $v_0$ and any composition $\x_0$, the process $\X_{\tau_v^K}^K$ is the stochastic composition of the CRN at the first time it reaches volume $v$. For the typical deterministic volume $V_t^K = v_t = e^{\Lambda t}v_0$, this gives
\begin{equation}
\mbb{P}_{\x_0,v_0}\!\left(\X_{t}^K\approx \x
     \right) 
    \asymp \exp\{-K v_0 e^{\Lambda t}\mcl{I}(\x) \},
    \label{eq:composition_ldp_time}    
\end{equation} 
provided that fluctuations in $V_t^K$ are sufficiently small. We term the non-negative function \(\mcl{I}\) the \textit{composition quasipotential}. It quantifies composition fluctuations away from $\x^\star$ among trajectories that have attained the same volume. 
Note that the effective large-deviation speed is not only the
system-scale parameter \(K\), but the unnormalized growing volume \(K v\). 

For all composition $\x \in \Delta_w$, conjugate momentum $\p \in T_\x^*\Delta_\w$ (a cotangent vector at $\x$) with $\p_D = 0$, and contact coordinate $u \in \Real$ we define the contact Hamiltonian
\begin{equation}
    \mbb{K}(\x,\p,u)
    =
    \sum_{r=1}^R
    \left[
    \exp\left\{
    \p^\top(\St^r-\,\nu_r\x)+\nu_r u
    \right\}
    -1
    \right]\sj_r(\x),
    \label{eq:contact_hamiltonian}
\end{equation}
Notice that $\mbb{K}$ is defined on a $2D-1$ dimensional domain. The dependence of $\mbb{K}$ on the value variable \(u\) makes it a contact Hamiltonian. The composition quasipotential $\mcl{I}$ is characterized by a contact
Hamilton--Jacobi equation. In the interior of the simplex, \(\mcl{I}\) satisfies the zero-contact-energy condition
\begin{equation}
    \mbb{K}(\x,\nabla \mcl{I},\mcl{I})=0, \qquad 
    \mcl{I}(\x^\star)=0.    
    \label{eq:contact_hj}
\end{equation}
 For dilute systems where \(\nu_r=0\), the \(u\)-dependence disappears, and the equation formally reduces to the usual Hamilton--Jacobi equation for a constant-volume CRN \cite{lazarescu2019large,gagrani2023action}.

\subsection{Moments of observables}
\label{sec:observable_moment_result}

Our second result concerns the long-time moments of the intensive real-time
observables introduced in Sec.~\ref{sec:observables}.  We collect these observables into the vector
\begin{equation}
    \wh{\mcl{O}}_t^K
    =
    \left(
    \widehat F_{1,t}^K,\ldots,\widehat F_{m,t}^K,
    \widehat G_{1,t}^K,\ldots,\widehat G_{\ell,t}^K
    \right)\in\Real^{m+\ell}.
    \label{eq:intensive_observable_vector_real_time}
\end{equation}
The conjugate bias vector is
\begin{equation}
    \theta
    =
    (\alpha,\beta)
    =
    (\alpha_1,\ldots,\alpha_m,\beta_1,\ldots,\beta_\ell)\in\Real^{m+\ell},
    \label{eq:real_time_bias_vector}
\end{equation}
where \(\alpha_i\) biases the state-time observable \(\wh{F}_i\), and
\(\beta_j\) biases the reaction-count observable \(\wh{G}_j\).  For each
reaction \(r\), we write
\(
    g_r
    =
    (g_{1,r},\ldots,g_{\ell,r})
\)
for the vector of reaction-observable increments associated with that
reaction. 

By transforming to the accumulated-volume clock, as derived in Appendix, the scaled-cumulant generating function (SCGF) in the long-time horizon is defined as
\begin{equation}
    \eta(\theta):=\lim_{a\to\infty}\frac{1}{a}\z_\theta(\x,v,a),
\end{equation}
where $\z_\theta$ satisfies the Hamilton-Jacobi equation
\begin{equation}
    \partial_a\z_\theta(\x,v,a)
    =
    -
    \widetilde H_\theta
    \left(
    \x,v;
    \nabla_\x\z_\theta,
    \partial_v\z_\theta
    \right),
    \nonumber
\end{equation}
with terminal condition
\begin{equation}
    \z_\theta(\x,v,a_f)=0,
    \;\;
    \nabla_\x\z_\theta(\x,v,a_f)=\0,
    \;\;
    \partial_v\z_\theta(\x,v,a_f)=0,
\end{equation}
and the tilted-Hamiltonian 
\begin{align*}
    &\widetilde H_\theta(\x,v;\pp,u)
    =
    \;
    \alpha\cdot f(\x)
    \nonumber
    \\
    +&
    \sum_{r=1}^R
    \left[
    \exp\left\{
    \frac{\pp^\top}{v}(\St^r-\,\nu_r\x)
    +
    \nu_r u
    +
    \beta\cdot g_r
    \right\}
    -
    1
    \right]\sj_r(\x).
\end{align*}
The long-time moments of the observables conditioned on systems that do not go extinct are obtained by differentiating the SCGF
\(\eta(\theta)\) at \(\theta=\0\). In particular, the mean of the observable $\wh{\mcl{O}}^K_{t,i}$ is
\begin{equation}
    \lim_{t\to\infty}
    \lim_{K\to\infty}
    \mbb{E}
    \left[
    \wh{\mcl{O}}_{t,i}^K
    \right]
    =
    \partial_{\theta_i} \eta(\theta)\big|_{\theta=\0},
    \label{eq:observable_mean_from_eta}
\end{equation}
and, for two components \(\wh{\mcl{O}}_{t,i}^K\) and \(\wh{\mcl{O}}_{t,j}^K\), their covariance is
\begin{equation}
    \lim_{t\to\infty}
    \lim_{K\to\infty}
    K a_t^* \,
    \mathrm{Cov}
    \left(
    \wh{\mcl{O}}_{t,i}^K,
    \wh{\mcl{O}}_{t,j}^K
    \right)
    =
    \partial_{\theta_i}\partial_{\theta_j}
    \eta(\theta)\big|_{\theta=\0},
    \label{eq:component_covariance_from_eta}
\end{equation}
where we introduce the deterministic accumulated-volume clock
\begin{equation}
    a_t^* = \int_0^t v_0 e^{\Lambda s}\,ds = \frac{e^{\Lambda t}-1}{\Lambda}v_0. \label{eq:det_integrated_volume}
\end{equation}
We remark that, for growing systems with \(\Lambda>0\), the covariance decays
exponentially in physical time,
\(
    \mathrm{Cov}\!\left(\wh{\mcl O}_t^K\right)
    \sim 
    O\!\left(e^{-\Lambda t}\right).
\)
The constant-volume result is recovered in the limit \(\Lambda\to0\), where the covariance has the usual long-time scaling 
\( \mathrm{Cov}\!\left(\wh{\mcl O}_t^K\right)
    \sim
    O(t^{-1})\) (see \cite{Touchette2009,lazarescu2019large}).

The exact tilted problem is a nonlinear contact eigenvalue problem,
\begin{equation}
    E_\theta
    =
    \mbb K_\theta
    \left(
        \x,
        \nabla_\x W_\theta(\x),
        W_\theta(\x)
    \right),
    \label{eq:contact_hj_eigenvalue_problem}
\end{equation}
with the tilted contact Hamiltonian
\begin{align}
    &\mbb K_\theta(\x,\p,u)
    =
    \alpha\cdot f(\x)+\nonumber\\
    &
    \sum_{r=1}^R
    \left[
        \exp\left\{
            \p^\top(\St^r-\nu_r\x)
            +
            \nu_r u
            +
            \beta\cdot g_r
        \right\}
        -
        1
    \right]\sj_r(\x).
   \nonumber
\end{align}
 This problem is difficult
because the eigenfunction enters not only through its gradient
\(\nabla_\x W_\theta\), but also through its value \(W_\theta\) itself,
reflecting the coupling between composition fluctuations and volume growth.
For this reason, we use a reduced approximation that removes this explicit
contact dependence and replaces the full contact eigenvalue problem by a
Hamiltonian problem on composition space.

We define the tilted reduced Hamiltonian
\begin{align}
    &\mbb{G}_\theta(\x,\p)
    :=
    \mbb K_\theta(\x,\p,0)
     \label{eq:tilted_reduced_hamiltonian} \\
    &= \alpha\cdot f(\x) +\sum_{r=1}^R
    \left[
    \exp\left\{
    \p^\top (\St^r-\,\nu_r\, \x)
    +
    \beta\cdot g_r
    \right\}
    -
    1
    \right]\sj_r(\x),\nonumber
\end{align}
where $x_D = 1- \sum_d^{D-1}x_d$ and $p_D = 0$. Note that $\mbb{G}$ is symplectic and defined on a $2D-2$ dimensional domain.
Let \((\x_\theta^\star,\p_\theta^\star)\) be the fixed point of the reduced
tilted Hamiltonian dynamics:
\begin{align}
    \frac{d \x_\theta}{dt}&=\phantom{-}\partial_\p \mbb{G}_\theta(\x_\theta^\star,\p_\theta^\star)
    =
    \0,\nonumber\\
    \frac{d \p_\theta}{dt}&=-\partial_\x \mbb{G}_\theta(\x_\theta^\star,\p_\theta^\star)
    =
    \0.
    \label{eq:tilted_fixed_point}
\end{align}
The value of the reduced tilted Hamiltonian at this fixed point is
\begin{equation}
    E_\theta
    =
    \mbb{G}_\theta(\x_\theta^\star,\p_\theta^\star).
    \label{eq:E_theta}
\end{equation}


Finally, we define the escape correction
\begin{equation}
    \epsilon_\theta
    =
    \int_{\gamma_\theta} \p\cdot d\x ,
    \label{eq:escape_correction}
\end{equation}
where \(\gamma_\theta\) denotes the reduced escape path satisfying Hamilton's
equations of motion from \((\x_\theta^\star,\p_\theta^\star)\) to the
terminal condition \(\p=\0\).  
Using these quantities, independently of the initial $(\x_0,v_0)$, the asymptotic (SCGF) can be approximately written as
\begin{equation}
    \eta(\theta)
    \simeq
    E_\theta-\Lambda \epsilon_\theta.
    \label{eq:eta_theta}
\end{equation}
Then, the mean and covariance-finding procedure reduces to computing
\(\eta(\theta)\) and differentiating it at zero bias.

\section{Applications}
\label{sec:applications}

\begin{figure*}[t]
    \centering
    \begin{subfigure}{.48\textwidth}
        \centering
        \includegraphics[width=\linewidth]{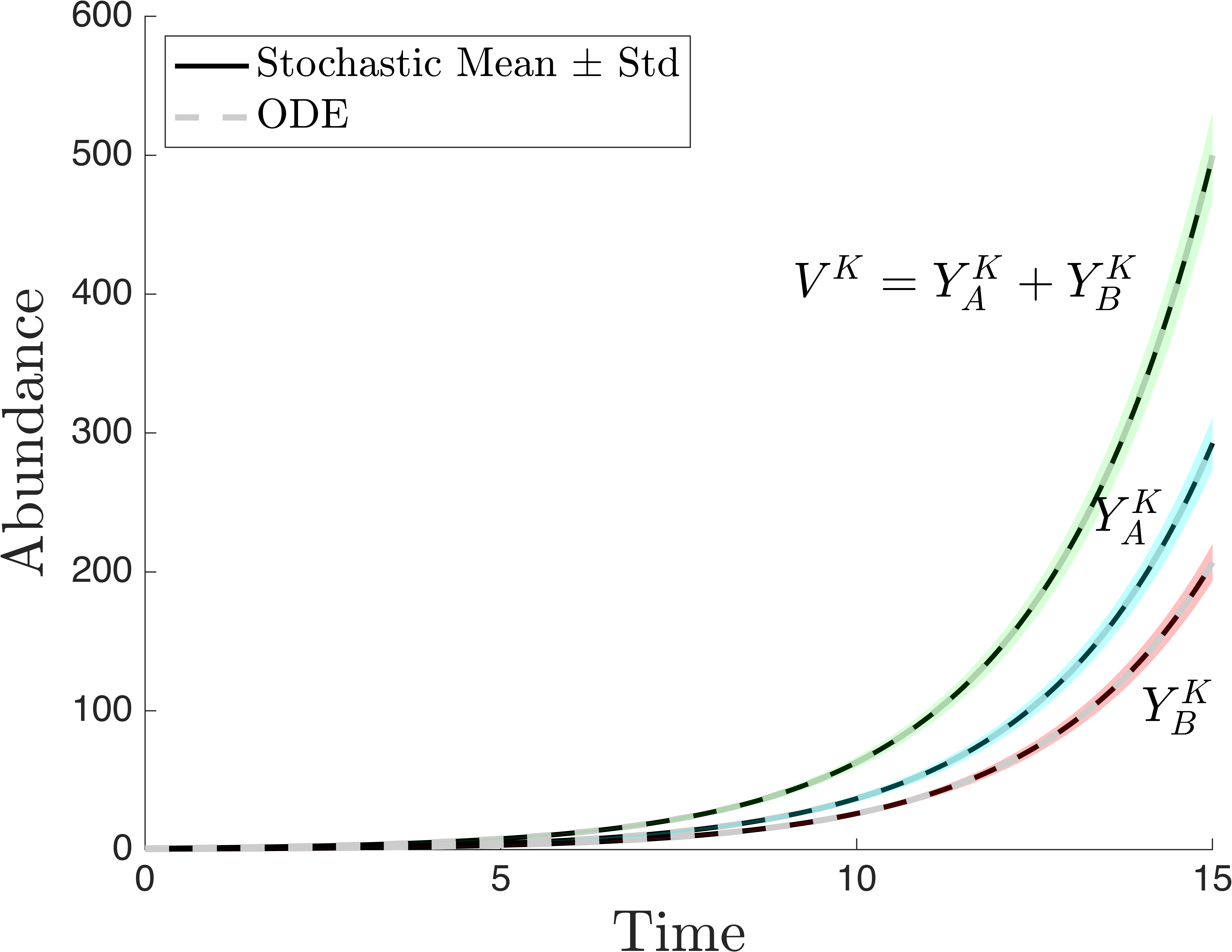}
    \end{subfigure}
    \hfill
    \begin{subfigure}{.48\textwidth}
        \centering
        \includegraphics[width=\linewidth]{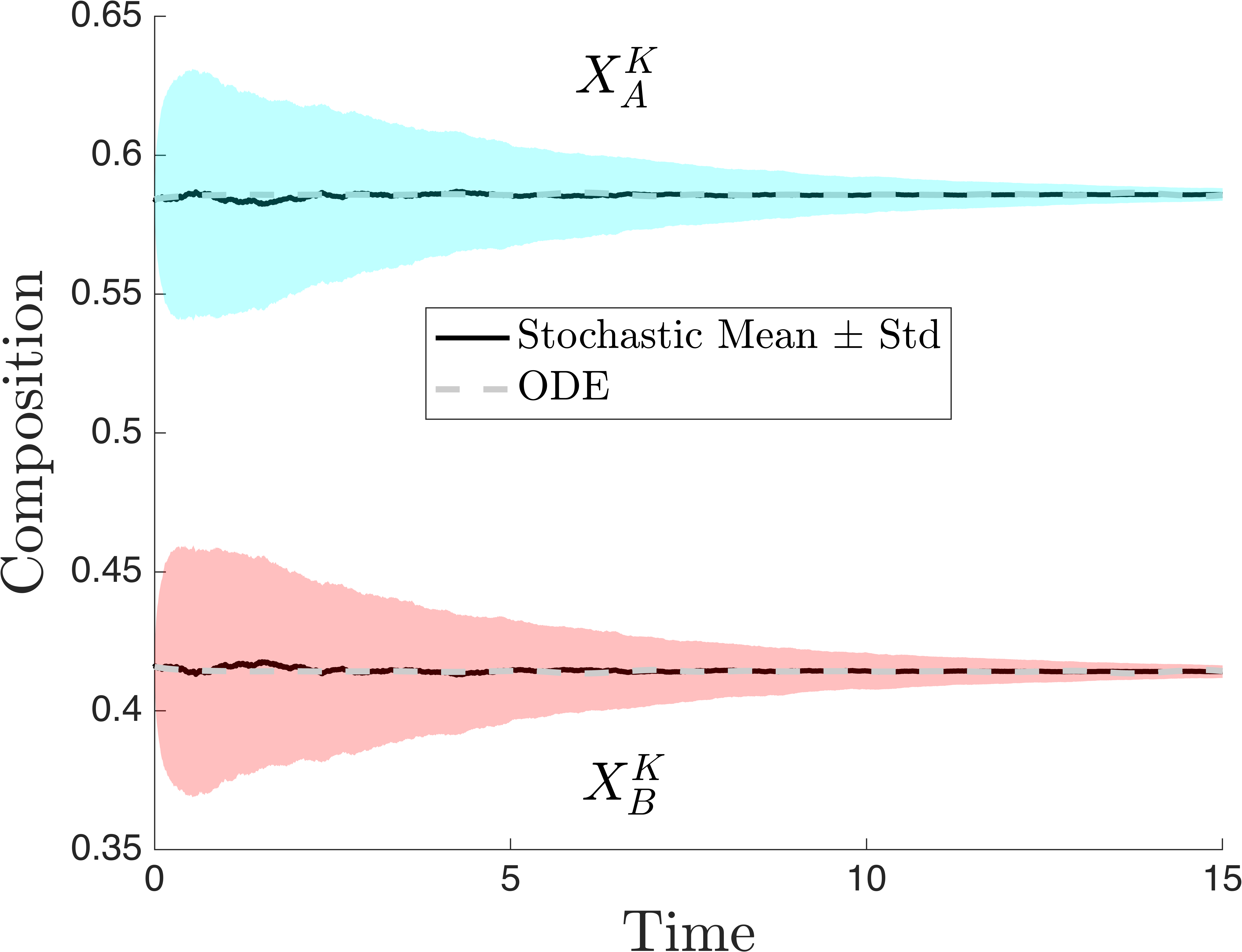}
    \end{subfigure}
    \caption{
    Abundance and composition dynamics for the two-species autocatalytic
    network.  We simulated \(2000\) stochastic trajectories with system-scale
    parameter \(K=101\), initialized at \((n_A,n_B)=(59,42)\).  The abundance
    variables and total volume grow exponentially, while the composition
    remains near the attracting balanced-growth composition.  The composition
    variance decreases as the volume grows.
    }
    \label{fig:AC_ode}
\end{figure*}

\begin{figure*}[t]
    \centering
    \begin{subfigure}{.48\textwidth}
        \centering
        \includegraphics[width=\linewidth]{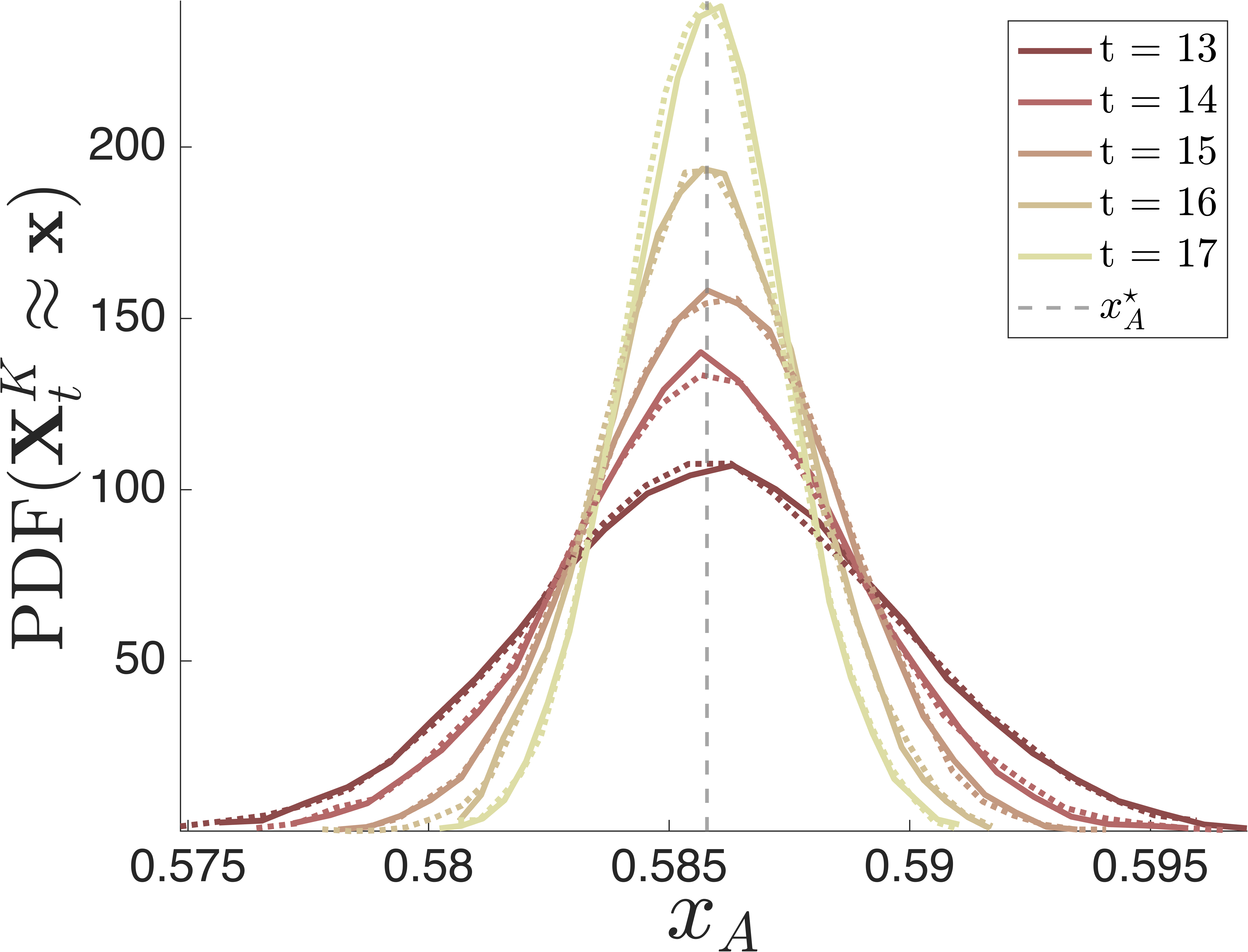}
    \end{subfigure}
    \hfill
    \begin{subfigure}{.48\textwidth}
        \centering
        \includegraphics[width=\linewidth]{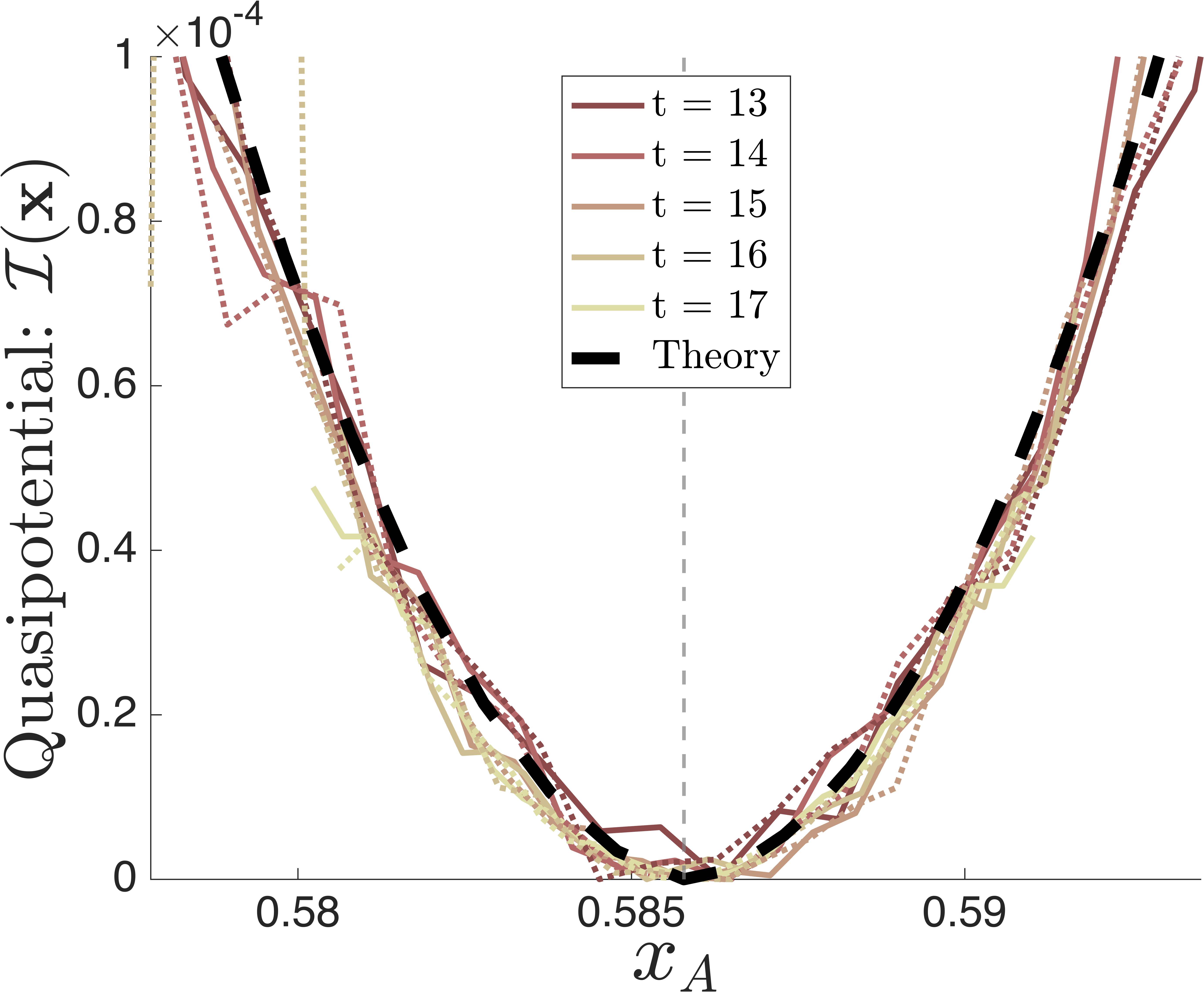}
    \end{subfigure}
    \caption{
    Composition distributions and quasipotential for the two-species
    autocatalytic network.  The left panel shows composition probability
    densities at increasing volumes or nominal balanced-growth times.  The
    right panel shows the empirical quasipotential,
    \(-\log(\mathrm{PDF})/(Kv)\), overlaid with the solution of
    Eq.~\eqref{eq:example_contact_hj}.  Solid curves correspond to
    fixed-volume sampling; dotted curves correspond to physical-time sampling.
    Their close agreement in this example indicates that volume fluctuations
    are subleading for the displayed compositions and times.
    }
    \label{fig:AC_quasipot}
\end{figure*}
As shown in Ref.~\cite{gagrani2026topological}, any chemical reaction network model of
cellular growth that supports exponential growth for a broad class of kinetics
must contain autocatalysis.  We therefore illustrate the theory on two
autocatalytic growth networks.  The first is a minimal two-species
autocatalytic network, which serves as a numerically tractable example
\cite{blokhuis2020universal,gagrani2024polyhedral}.  The second is a
coarse-grained model of cellular growth with transport machinery, precursor,
and ribosomes, motivated by resource-allocation models of bacterial growth
\cite{giordano2016dynamical,pandey2016analytic}.

Both networks have parameter regimes of exponential growth, and both exhibit the two scaling predictions
of the theory.  First, composition fluctuations collapse with large-deviation
speed equal to the system volume \(K v\), where \(K\) is the system-scale parameter and \(v\) is the
current normalized volume.  Second, covariance of accumulated-volume observables decays as \((K a_t^*)^{-1}\) with a proportionality factor that is numerically estimated from theory.  We also compare
accumulated-volume estimators with conventional time-averaged estimators. We return in Sec.~\ref{sec:discussion} to how our framework can be used in
experimental settings.

\begin{figure*}[t]
    \includegraphics[width=0.7\textwidth]{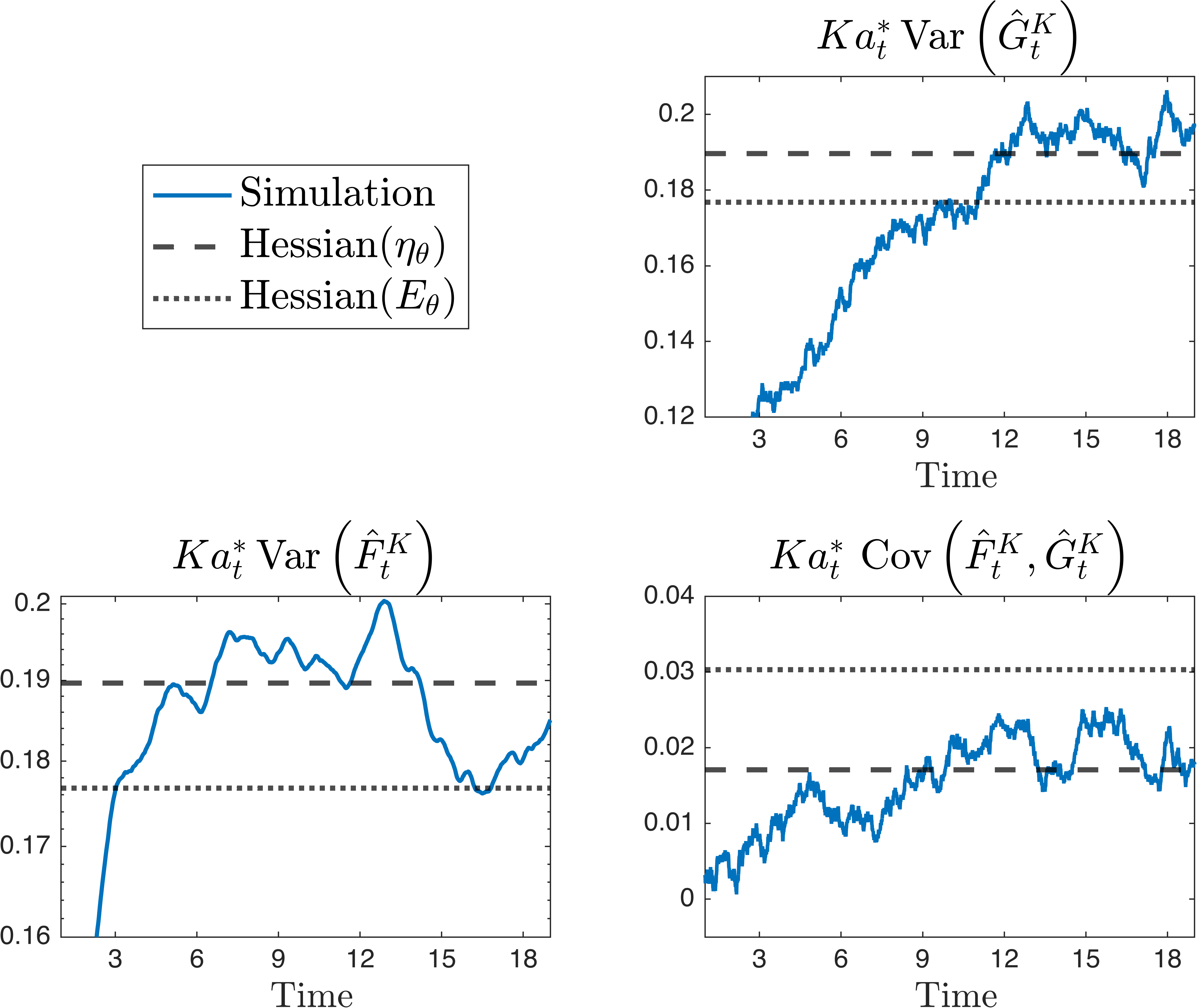}
    \caption{
    Variances and covariance of the accumulated-volume-averaged
    observables in Eq.~\eqref{eq:example_observables}. The variance and covariance at each time $t$ are scaled by the deterministic accumulated-volume level
    \(a_t^\star=v_0(e^{\Lambda t}-1)/\Lambda\). The theoretical
    predictions are obtained from the Hessian of the SCGF
    \(\eta_\theta\). An approximation using the Hessian of the fixed point energy \(E_\theta\) is also shown.  
    }
    \label{fig:AC_moments_descaled}
\end{figure*}

\begin{figure*}[]
    \centering
    \includegraphics[width=0.9\linewidth]{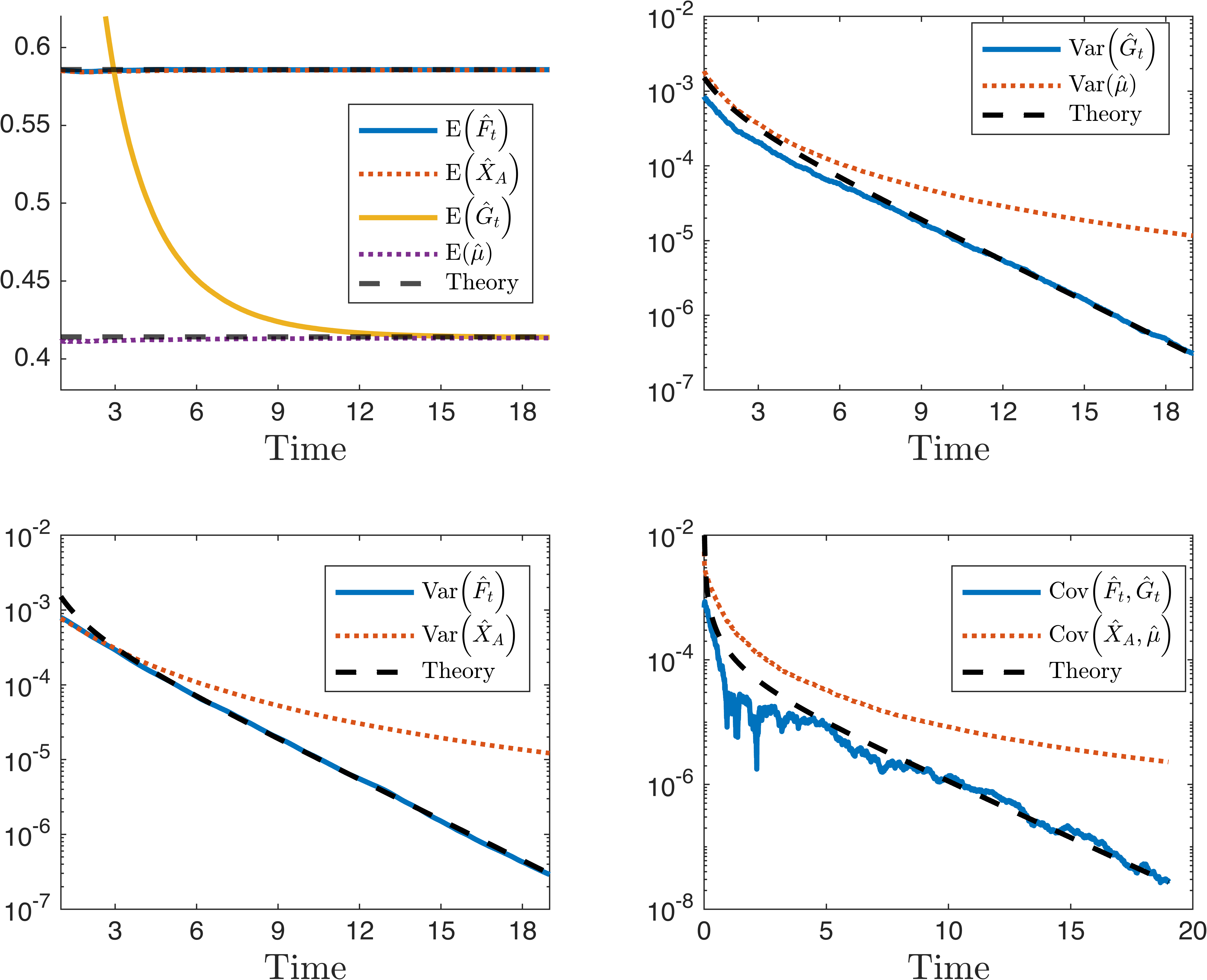}
    \caption{
    Fluctuations of composition and growth-rate estimators in the two-species
    autocatalytic network.  Accumulated-volume-averaged estimators are compared
    with the theoretical covariance prediction and with the corresponding
    conventional time-averaged estimators.  Both averaging schemes converge to
    the same balanced-growth means, but their fluctuations decay differently:
    the accumulated-volume estimators decay on the accumulated-volume scale,
    whereas the time-averaged estimators decay $O(t^{-1})$ in physical time.
    For the parameters shown, the accumulated-volume estimators consequently
    attain smaller variances at long times.
    }
    \label{fig:2AC_moment}
\end{figure*}

\subsection{A minimal autocatalytic network}
\label{sec:application_two_species}

We first consider the network
\begin{equation}
    \left\{\ce{A ->[1] B}, \;\; \ce{B ->[1] 2 A}\right\}.
\end{equation}
With species ordered as
\((A,B)\), the stoichiometric matrix is
\[
    \St
    =
    \begin{bmatrix}
        -1 & 2\\
         1 & -1
    \end{bmatrix},
\]
and the mass-action propensities are \(\J_1(\n)=n_A\) and
\(\J_2(\n)=n_B\).  Hence the deterministic rates are
\(\jj_1(\y)=y_A\) and \(\jj_2(\y)=y_B\), giving
\[
    \dot y_A=-y_A+2y_B,
    \qquad
    \dot y_B=y_A-y_B .
\]
We choose \(\w=(1,1)^\top\), so that \(v=y_A+y_B\), \(\x=\y/v\), and
\(x_A+x_B=1\).  The volume increments are \(\nu_1=0\) and \(\nu_2=1\), so
only the autocatalytic reaction \(\ce{B -> 2A}\) changes the volume.  Since
\(j_r(v\x)=v\sj_r(\x)\), the composition-level rates are
\(\jjj_1(\x)=x_A\) and \(\jjj_2(\x)=x_B\).

The composition--volume dynamics are therefore
\begin{align}
    \dot x_A&=-x_A+2x_B-x_Ax_B,
    \qquad
    \dot x_B=x_A-x_B-x_B^2,\nonumber\\
    \dot v &=x_Bv .
\end{align}
The composition dynamics have a unique stable balanced-growth fixed point
\[
    x_A^\star=2-\sqrt2
    =
    \frac{\sqrt2}{1+\sqrt2}
    \approx0.586,
    \qquad
    x_B^\star=\sqrt2-1\approx0.414 .
\]
The corresponding exponential growth rate is
\(\Lambda=x_B^\star=\sqrt2-1\).  The abundance and composition trajectories are shown in
Figure~\ref{fig:AC_ode}.  The abundance variables grow exponentially, while the
composition variables concentrate near \(\x^\star\).

\subsubsection{Composition quasipotential}
\label{subsec:application_ac_quasipotential}

Since \(x_A+x_B=1\), there is one independent composition coordinate.  We
write \(x=x_A\) and \(x_B=1-x\).  The contact Hamiltonian (Eq.~\ref{eq:contact_hamiltonian}) becomes
\begin{equation*}
    \mbb K(x,p,u)
    =
    x(e^{-p}-1)
    +
    (1-x)
    \left(
        e^{(2-x)p+u}-1
    \right),
\end{equation*}
where \(p\) is conjugate to \(x\), and \(u\) is the contact coordinate.  The
composition quasipotential satisfies the zero-contact-energy equation
\begin{equation}
    \mbb K(x,\mcl I'(x),\mcl I(x))=0,
    \qquad
    \mcl I(x^\star)=0.
    \nonumber
    \label{eq:example_contact_hj}
\end{equation}
Equivalently, along a zero-energy characteristic,
\[
    \dot x=\partial_p\mbb K,
    \qquad
    \dot p=-\partial_x\mbb K-p\,\partial_u\mbb K,
    \qquad
    d\mcl I=p\,dx .
\]
Using $\mcl{I}(x) = \int_{x^*}^x p(z)\,dz$,
\(p(x)\) is determined implicitly by
\begin{equation}
    \mbb{K}\pars{
    x,
    p(x),
    \int_{x^\star}^{x} p(z)\,dz
    }
    =
    0.
    \nonumber
\end{equation}
Numerically, starting from \(p(x^\star)=0\), one may solve this equation iteratively on a
grid.  For a small step \(\Delta x\), the first step is obtained from
\begin{equation}
    \mcl{I}(x^\star\pm \Delta x)
    \approx
    \pm p(x^\star\pm \Delta x)\Delta x. \nonumber
\end{equation}
Repeating this procedure gives \(p(x^\star\pm i\Delta x)\) and hence the full
branch \(p(x)\).  

We compare the quasipotential with stochastic simulations in
Figure~\ref{fig:AC_quasipot}. The probability distributions of compositions at a fixed volume are shown in the left panel. Observe that, as the volume increases, the distribution concentrates around the fixed composition attractor. This is expected as, in the limit $v\to\infty$, the system must become deterministic and the distribution collapses at the composition attractor. 
For fixed target volume \(v\), Eq.~\ref{eq:composition_ldp}
predicts
\(
  -
    \log
    \mbb P_{v_0}\!\left(
        X_{\tau_v^K}^K\approx x        
     \,\middle|\,
        \tau_v^K <\infty
        \right)/(Kv) \simeq
    \mcl I(x)
\)
as $Kv \to \infty$. Since all reactions have a nonnegative contribution to the volume, the system in the exponentially growing regime will reach $v$ in finite time almost surely, and the conditioning can be ignored for this model. 
Accordingly, the empirical quantity
\(
  -
    \log
    \mbb P_{v_0}\!\left(
        X_{\tau_v^K}^K\approx x        
    \right)/(Kv)
\)
should collapse onto \(\mcl I(x)\) as \(Kv\) grows.
This is verified in the right panel of Figure \ref{fig:AC_quasipot}.

In experiments and simulations, one often samples at prescribed physical
times rather than at fixed volumes. These are shown in dotted lines in the same figure.  If \(v_t=v_0e^{\Lambda t}\) is the
deterministic balanced-growth volume, then physical-time histograms are
expected to have the same leading profile,
\(
  -
    \log
    \mbb P_{v_0}\!\left(
        X_{\tau_v^K}^K\approx x        
    \right)/(Kv)
      \simeq
    \mcl I(x)
\)
when the volume distribution is sufficiently concentrated near \(v_t\) over
the range of compositions being probed.  Appendix
derives the corresponding volume-averaging approximation and gives the first
correction.  This correction explains why the agreement between fixed-volume
and fixed-time profiles is strongest near the balanced-growth composition.

\subsubsection{Moments}
\label{subsec:example_moments}

We next test the moment theory using one state-dependent and one jump-dependent
observable.  We take \(f(\x)=x_A\) for the composition and
\(g_r=\nu_r\) for the growth rate.  The corresponding \(F\)- and \(G\)-type
estimators introduced in Sec.~\ref{sec:observables} are therefore
\begin{equation}
    \widehat F_t^K
    =
    \frac{\int_0^t V_s^K X_{A,s}^K\,ds}{A_t^K},
    \qquad
    \widehat G_t^K
    =
    \frac{\nu^\top\Om_t^K}{K A_t^K}
    =
    \frac{V_t^K-V_0^K}{A_t^K}.
    \label{eq:example_observables}
\end{equation}
Their limiting means are \(x_A^\star\) and \(\Lambda\), respectively.

Writing \(x=x_A\), \(x_B=1-x\), and introducing biases
\(\theta=(\alpha,\beta)\) conjugate to
\((\widehat F_t^K,\widehat G_t^K)\), the reduced tilted Hamiltonian is
\begin{equation}
    \mbb G_\theta(x,p)
    =
    x\left(e^{-p}-1\right)
    +(1-x)\left(e^{(2-x)p+\beta}-1\right)
    +\alpha x .
    \nonumber
\end{equation}
For each small bias \(\theta\), we numerically determine the tilted stationary
point and the escape trajectory to the terminal manifold \(p=0\).  The latter
is obtained by shooting from the unstable direction of the tilted stationary
point.  From the stationary value \(E_\theta\) and escape action
\(
    \epsilon_\theta
    =
    \int_{\gamma_\theta}p\,dx ,
\)
we construct
\begin{equation}
    \eta(\theta)
    =
    E_\theta-\Lambda\epsilon_\theta ,
    \label{eq:example_eta_theta}
\end{equation}
and evaluate its Hessian at zero bias by finite differences.

The theory then predicts
\begin{equation}
    \lim_{t\to\infty}
    \lim_{K\to\infty}
    K a_t^\star\,
    \mathrm{Cov}
    \begin{pmatrix}
        \widehat F_t^K\\
        \widehat G_t^K
    \end{pmatrix}
    =
    \nabla_\theta^2\eta(\0),
    \qquad
    a_t^\star
    =
    v_0\frac{e^{\Lambda t}-1}{\Lambda}.
    \label{eq:example_covariance_prediction}
\end{equation}
As shown in Fig.~\ref{fig:AC_moments_descaled}, the rescaled simulation
covariances approach the predicted Hessian at long times.  We also show the
approximation obtained by omitting the escape correction,
\(\nabla_\theta^2E_\theta|_{\theta=\0}\).  For this example it is already
close to the full prediction, indicating that the escape correction is small
but nonzero.

Figure~\ref{fig:2AC_moment} shows the same fluctuations without rescaling and
compares them with the conventional time-averaged estimators defined in
Sec.~\ref{sec:observables}.  Both averaging schemes approach the same
balanced-growth means, but their variances have different asymptotic
behavior.  The accumulated-volume estimators weight later times more strongly,
when the system is larger and more reaction events are observed, and their
covariances decay on the scale \(1/(K A_t)\).  In the simulations this leads
to substantially smaller long-time fluctuations than for the corresponding
time-averaged estimators.

\subsection{A coarse-grained cellular growth model}
\label{sec:application_tpr}

\begin{figure}[t]
    \centering
    \includegraphics[width=0.9\linewidth]{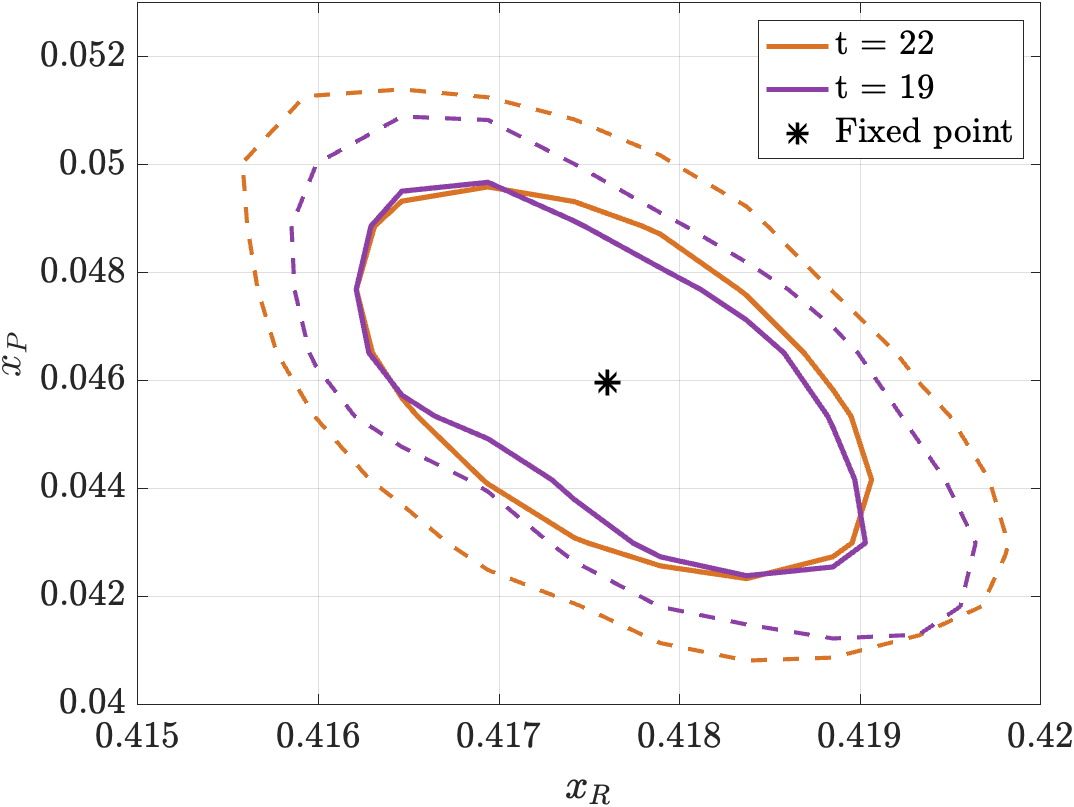}
    \caption{
    Empirical quasipotential collapse for the TPR model.  Curves show
    \(-\log\mbb P(\X_t^K\approx\x)/(K V_t)\) at several volumes or nominal
    balanced-growth times.  Collapse under the \(K V_t\) scaling supports the
    predicted composition-level large-deviation principle.
    We simulated 1000 stochastic trajectories with system-scale parameter $K=100$, initialized at $(n_T,n_P,n_R)=(58,5,42)$.
    }
    \label{fig:tpr_quasipotential}
\end{figure}

\begin{figure*}[t]
    \centering
    \includegraphics[width=0.9\linewidth]{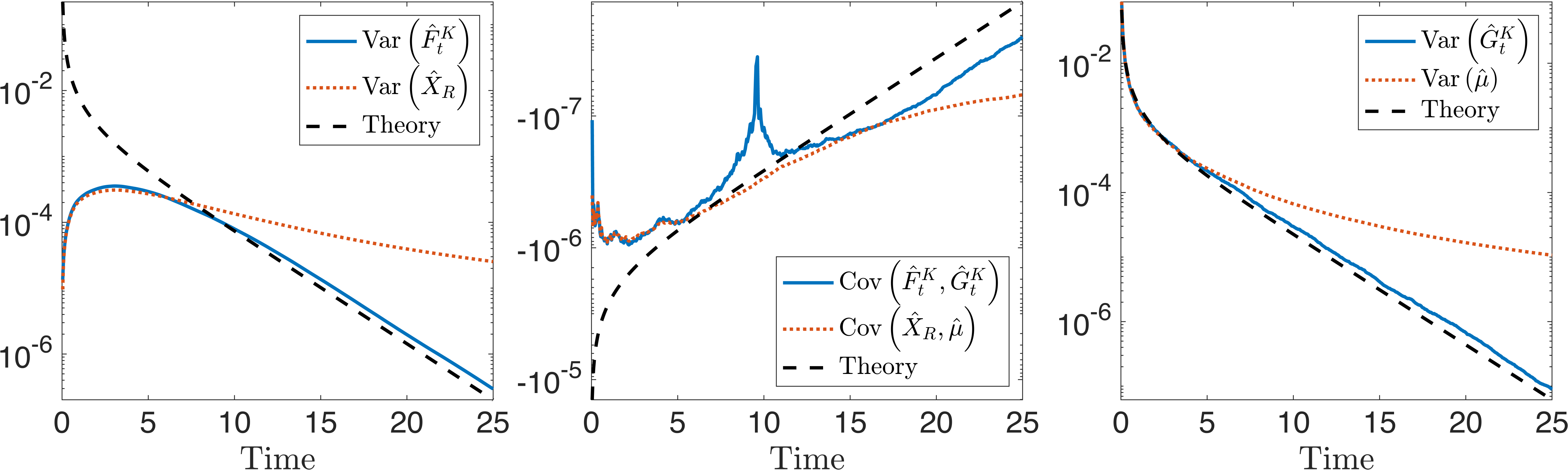}
    \caption{
Covariance scaling for ribosome-composition and growth-rate estimators in
the TPR model.  Accumulated-volume-averaged estimators show the predicted decay
with accumulated volume and agree closely with the theoretical covariance
calculation.  The remaining discrepancy is consistent with finite-time
relaxation effects and the small-bias heuristic.  conventional
time-averaged estimators converge to larger variances, reflecting the fact
that they weight early and late physical times equally rather than by reaction
activity.  The ribosome-composition--growth-rate correlation approaches a
negative value.
}
    \label{fig:TPR_moment}
\end{figure*}

The formalism developed above is naturally suited to chemical reaction network models of
cellular growth.  We illustrate this using 
the resource-allocation model of
Giordano et al.~\cite{giordano2016dynamical}.
The model describes a growing
cell in terms of three coarse-grained molecular species:
\begin{align*}
    T &\quad \text{transport/metabolic machinery},\\
    P &\quad \text{precursor pool},\\
    R &\quad \text{ribosomes}.
\end{align*}
The transport machinery converts external nutrient into internal precursor,
while ribosomes consume precursor to synthesize either more transport
machinery or more ribosomes.

We write the stochastic state as
\[
    \N_t^K
    =
    \left(
    N_{T,t}^K,
    N_{P,t}^K,
    N_{R,t}^K
    \right),
\]
and consider the reaction network
\begin{equation}
\{ T \longrightarrow T+P, \; \; P+R \longrightarrow T + R, \; \;P+R \longrightarrow 2 R\}     .\nonumber
\label{eq:tpr_reaction_ribosome}
\end{equation}
The first reaction represents precursor production by transport/metabolic
machinery.  The second and third reactions represent ribosome-catalyzed
synthesis of transport machinery and ribosomes, respectively.  Abundances have been normalized such that it takes $1$ unit of precursors for producing one unit of \(T\) or
\(R\).

Let
\[
    \Y_t^K=\frac{\N_t^K}{K},
    \qquad
    V_t^K=\w^\top\Y_t^K,
    \qquad
    \X_t^K=\frac{\Y_t^K}{V_t^K}.
\]
In deterministic variables, \(\y=V\x\), where $\y$ and $\x$ are, respectively, $\y = (y_T,y_P,y_R)$ and $\x=(x_T,x_P,x_R)$.
We use Michaelis--Menten kinetics for precursor production and translation.
For stochastic simulations, we use Gillespie dynamics with propensities
\begin{align*}
    \J_1^K(\n)
    &=
    K V^K(\n)\,
    \phi\, X_T^K(\n),
    \\
    \J_2^K(\n)
    &=
    K V^K(\n)\,
    (1-\alpha)\kappa_R X_R^K(\n)
    \frac{X_P^K(\n)}{\kappa_0+X_P^K(\n)},
    \\
    \J_3^K(\n)
    &=
    K V^K(\n)\,
    \alpha \kappa_R X_R^K(\n)
    \frac{X_P^K(\n)}{\kappa_0+X_P^K(\n)}.
\end{align*}
The deterministic rates are
\begin{align*}
    \jj_1(\y)
    &=
    \phi\, y_T,
    \\
    \jj_2(\y)
    &=
    (1-\alpha)\kappa_R 
    \frac{y_R y_P/V}{\kappa_0+y_P/V},
    \\
    \jj_3(\y)
    &=
    \alpha\kappa_R 
    \frac{y_R y_P/V}{\kappa_0+y_P/V}.
\end{align*}
Equivalently, using \(\y=V\x\), we write 
\(
\sj_r(\x) = j_r(V\x)/V,
\)
with
\begin{align*}
    \jjj_1(\x)
    &=
    \phi\, x_T
    \\
    \jjj_2(\x)
    &=
    (1-\alpha)\kappa_R \frac{x_R x_P}{\kappa + x_P},
    \\
    \jjj_3(\x)
    &=
    \alpha\kappa_R \frac{x_R x_P}{\kappa + x_P}.
\end{align*}
Thus, the model satisfies the scalability condition required by the theory.

In accordance with the model studied in \cite{giordano2016dynamical}, we choose the rate constants $(\kappa_R=1, \phi=0.704, \kappa_0=2.23)$ (corresponding to $eM = 1.57$ in Figure 3, \cite{giordano2016dynamical}). The volume contributions are taken to be $\w = (1,0,1)^\top$, and the associated volume change vector $\nu$ is:
\begin{equation}
    \nu_1=0,
    \qquad
    \nu_2=1,
    \qquad
    \nu_3=1.
    \label{eq:tpr_volume_changes}\nonumber
\end{equation}
The allocation parameter \(\alpha\in[0,1]\) controls
the fraction of ribosomal synthesis directed toward ribosome production.
For this model, the value of the allocation parameter that maximizes the growth rate is $\alpha^*=0.4176$. The composition attractor at $\alpha^*$ is $(x_T,x_P,x_R)=(0.58,0.05,0.42)$
with growth rate $\Lambda = 0.39$.

Ignoring the conditioning $\tau_V^K<\infty$, the theory predicts a composition large-deviation form
\(
    \mbb P_{v_0}(\X_t^K\approx\x)
    \asymp
    \exp[-K v \mcl I(\x)]   . 
\)
For this higher-dimensional model, we do not solve the contact
Hamilton--Jacobi equation for \(\mcl I\) explicitly.  Instead, using $v_t =v_0 e^{\Lambda t},$ we test the
predicted speed \(K v_t\).  Figure~\ref{fig:tpr_quasipotential} shows that
the empirical profiles collapse when plotted on this scale, supporting the
composition-level LDP.

We next consider two experimentally natural observables: the ribosomal
fraction \(x_R\) and the composition-level growth rate
\(\mu(\x)=\sum_{r=1}^3\nu_r\sj_r(\x)\).  As defined in Section~\ref{sec:observables}, we consider the
accumulated-volume-averaged estimators \(\wh{F}_t^K, \wh{G}_t^K\) and this conventional time averaged counterparts \(\wh{X}_R,\wh{\mu}\).

Figure \ref{fig:TPR_moment} shows the variances and covariances of the two sets of estimators. It can be seen that the covariance of the accumulated-volume-averaged estimators decays with $e^{-\Lambda t}$. Furthermore, the decay converges to the theoretical estimates calculated using the Hessian of the SCGF. To calculate the escape curves, we use a shooting method until the Hamiltonian trajectory reaches the zero-momentum manifold, \(\mathbf p=(0,0)\). Finally, similar to the previous example, 
time-averaged estimators decay at a slower rate, $O(t^{-1})$, compared to their accumulated-volume-average counterparts. 


\section{Discussion}
\label{sec:discussion}

In this work, we developed a large-deviation theory for stochastic CRNs that
grow exponentially.  Our approach exploits the variational structure of large
deviations through Hamilton--Jacobi theory.  Although the dynamics are
nonstationary in abundance space, their fluctuations acquire a simpler
structure after decomposing the state into an extensive volume coordinate and
an intensive, or ``co-moving,'' composition coordinate.  Our central results
identify two natural fluctuation scales.  Composition fluctuations are
controlled by the current unnormalized volume \(Kv_t\), whereas the covariances
of accumulated-volume-averaged observables decay with the deterministic
accumulated volume \(K a_t^\star\).  In addition to identifying these
large-deviation speeds, we characterize the corresponding rate functions and
provide numerical procedures for computing them.

An important technical point arises because growth need not be monotone in a
general CRN.  If some reactions decrease the volume, or if extinction is
possible, a trajectory need not reach an arbitrarily large target volume.
The natural object is therefore the composition at the first time
\(\tau_v^K\) that the stochastic volume reaches a prescribed value \(v\) in finite time.  Similarly, for observables
measured in accumulated-volume time, we condition on trajectories that survive
long enough for the accumulated volume to reach the prescribed level.  In the
examples studied here, the relevant volume increments are nonnegative, so
this conditioning is trivial.  In the general theory, however, it is
essential and allows the framework to accommodate reverse reactions,
degradation, death, and other mechanisms that can reduce volume or terminate
growth.

The framework has several implications for the analysis of data from growing
systems.  The first concerns composition fluctuations.  The theory predicts
that composition distributions satisfy a large-deviation principle with
speed \(Kv\) and rate function given by the composition quasipotential.  For
time-series abundance data, one may therefore first decompose each trajectory
into its composition and volume coordinates and then compare compositions at
a common first-attained volume.  If the observed stochasticity is generated
solely by the underlying stochastic CRN, the negative logarithms of these
composition distributions, normalized by the attained volume, should collapse
onto a common curve in the large-system and large-volume regime.  Within the
theory, this limiting curve is the composition quasipotential.

The second consequence is a fluctuation theory for accumulated observables.
The moment theory identifies accumulated volume as the natural clock for
reaction statistics and, correspondingly, accumulated-volume averages as
natural estimators for growing systems.  Because reaction activity increases
with system size, equal increments of accumulated volume---rather than equal
intervals of physical time---contain comparable amounts of reaction activity.
Conditioning on trajectories that reach the required accumulated-volume
level, the theory predicts that the covariances of these estimators decay as
\(O((K a_t^\star)^{-1})\).  

The covariance predictions also have a direct interpretation for experimental
data.  For two observables \(\wh O_t\) and \(\wh O'_t\), their correlation is
\[
    \rho
    =
    \frac{
        \operatorname{Cov}(\wh O_t,\wh O'_t)
    }{
        \sqrt{
            \operatorname{Var}(\wh O_t)
            \operatorname{Var}(\wh O'_t)
        }
    }.
\]
Because the variances and covariances have the same accumulated-volume
scaling, their ratio can converge to a finite value determined by the Hessian
of the SCGF.  This is particularly relevant for interpreting single-cell
growth data, where experiments increasingly resolve not only mean
compositions and growth rates but also correlations across cells within the
same growth condition.  Given a stochastic CRN model for the underlying
metabolism, our theory provides a baseline expectation for correlations
arising from intrinsic reaction noise alone, before invoking additional
cellular mechanisms or extrinsic sources of variability.

Our second application illustrates this point using the resource-allocation
model of Ref.~\cite{giordano2016dynamical}, whose parameters were fitted to
reproduce growth-law behavior across growth conditions.  In
Ref.~\cite{pavlou2025single}, ribosomal investment and growth rate are found
to be strongly correlated across different growth media, while their
single-cell association within a fixed medium is considerably weaker.  This
observation is not inconsistent with the present example.  In Fig.~\ref{fig:TPR_moment},
the coarse-grained model predicts a negative within-population correlation
between ribosomal composition and growth-rate fluctuations.  We do not
interpret this result as a quantitative fit to the experimental data.
Rather, it illustrates how the framework can be used to formulate and test
mechanistic null hypotheses for single-cell covariance structure and,
potentially, to identify which additional mechanisms are required by the
data.

Several directions remain open.  First, the present work focuses primarily on
systems with a stable balanced-growth composition.  Systems with multiple
balanced-growth states would require a global quasipotential constructed from
least-action transitions between competing growth modes
\cite{gagrani2023action}.  Second, although genetically engineered cells can
in some cases be maintained for prolonged periods in a balanced-growth
regime, natural cells generally undergo repeated cycles of growth and
division.  Incorporating division would connect the present framework to
lineage and population statistics and could provide more systematic null
models for single-cell variability
\cite{pandey2020exponential,pandey2025division}.  Third, our moment calculation uses a symplectic reduction of an underlying
contact eigenvalue problem.  To our knowledge, the variational and geometric
structure of such contact eigenvalue problems has not been systematically
developed in the large-deviation setting.  A more complete treatment would
clarify the role of the contact eigenvectors, the accuracy of the reduced
approximation used here, and the construction of driven processes that
realize atypical fluctuations in supercritical growing systems. 
Finally, extending the theory to dilution, environmental fluctuations, and
interacting populations would make it more directly applicable to microbial
growth, metabolic ecology, and experimental studies of cellular
stochasticity.

\section*{Acknowledgements}
PG acknowledges Atsushi Kamimura and Artemy Kolchinsky for several helpful discussions. 
This research is supported by JSPS Grant in Aid for Scientific Research (B) (24K00542), JSPS Grant-in-Aid for Transformative Research Areas (A) (25H01365) and  
JST CREST (JPMJCR25Q2). The authors used ChatGPT (OpenAI) to assist with language editing and organization of the manuscript. 

\bibliography{bibliography}

\end{document}